\documentclass[11pt]{article}
\usepackage{moriond,epsfig}
\usepackage{xspace}
\usepackage{multicol}
\usepackage{calc}
\usepackage[compact]{titlesec}

\newcommand{\un}[1]{\ensuremath{\, \mathrm{#1}}}

\def\journal#1#2#3#4{{#1} {\bf #2}, #3 (#4)}
\def\PRL{\em Phys. Rev. Lett.}
\def\PRD{{\em Phys. Rev.} D}
\def\PR{\em Phys. Rev.}
\def\ApP{\em Astrop. Ph.}
\def\ApJ{\em Astrop. J.}
\def\AA{\em Astron. \& Astroph.}
\def\NIMA{{\em NIM} A}
\def\NPB{{\em Nuc. Ph.} B}
\def\nat{\em Nature}

\newlength{\delimw}
\newlength{\rxfc} 

\newcounter{temp}

\DeclareRobustCommand{\deg}{\ensuremath{^{\circ}}}

\newcommand{\ama}{\textsc{AMANDA}\xspace}

\newcommand{\icecube}{\textsc{IceCube}\xspace}
\newcommand{\deepcore}{\textsc{DeepCore}\xspace}
\newcommand{\inice}{\textsc{InIce}\xspace}
\newcommand{\icetop}{\textsc{IceTop}\xspace}

\newcommand{\wimp}{\textsc{WIMP}\xspace}

\newcommand{\magic}{\textsc{Magic}\xspace}
\newcommand{\veritas}{\textsc{Veritas}\xspace}

\newcommand{\swift}{\textsc{Swift}\xspace}

\newlength{\figwidth}
\newlength{\figsep}
\newlength{\pba} \newlength{\wa} 
\newlength{\pbb} \newlength{\wb} 
\newlength{\pbc} \newlength{\wc} 
\newlength{\ha} 

\newcommand{\twofig}[5][tbp]{
  \settowidth{\wa}{\includegraphics{#2}}
  \settoheight{\ha}{\includegraphics{#2}}
  \settowidth{\wb}{\includegraphics[height=\ha]{#4}}
  \setlength{\pba}{(\textwidth-\figsep)*\ratio{\wa}{\wa+\wb}}
  \setlength{\pbb}{(\textwidth-\figsep)*\ratio{\wb}{\wa+\wb}}
  \begin{figure}[#1]
    \parbox[t]{\pba}{\includegraphics[clip=true, width=\pba]{#2}\caption{#3}}
    \hspace*{-.5\pba}\hfill
    \parbox[t]{\pbb}{\includegraphics[clip=true, width=\pbb]{#4}\caption{#5}}
  \end{figure}
}
  
\newlength{\thinlinewidth} 
\newlength{\thicklinewidth} 

\begin{document}
\vspace*{4cm}
\title{RECENT RESULTS FROM ICECUBE ON NEUTRINOS AND COSMIC RAYS}

\author{ S.B\"OSER for the \icecube collaboration~\footnote{http://www.icecube.wisc.edu} }

\address{Physikalisches Institut, Universit\"at Bonn, Nu\ss{}allee 12\\
53115 Bonn, Germany}

\maketitle\abstracts{Encompasing a volume of $\sim1\un{km^3}$ of glacial ice at
the South Pole, \icecube is
currently the worlds largest neutrino detector.
It consists of 5160 optical modules on 86 strings in a depth
between 1450\un{m} and 2450\un{m}, as well as 324 optical modules
arranged in 81 stations on the surface to detect charged cosmic rays.
A large amount of data has already been acquired with smaller configurations throughout the
installation period.\\
Using this data the atmospheric neutrino spectrum in the northern hemisphere has
been measured up to 100\un{TeV}. No point sources have been identified in a set
of more than $10^5$ neutrino candidates from both hemispheres. Searches for transient sources have set
stringent limits on neutrino emission from gamma-ray bursts, and are now
accompanied by an extensive neutrino-triggered follow-up program.
A very large statistics of cosmic ray events has revealed an anisotropy in the
cosmic ray flux on the $10^{-3}$ level in the $10-100\un{TeV}$ range. While no
sources of extra-terrestrial neutrinos have been found yet, the physics results
obtained so far illustrate the very good performance of the detector.}

\section{Introduction}
The \icecube detector depicted in Figure~\ref{fig:icecube} employs the $\sim3\un{km}$ thick glacial ice cap at the
South Pole as a target for
neutrino-nucleon interactions. The emerging high-energetic charged particles
emit Cherenkov light that is detected by optical modules embedded in the
ice. Each optical module consists of a photomultiplier tube in a pressure
housing~\cite{dompaper}. In its now final configuration a total of 86 strings
holding 60 modules each has been deployed throughout seven consecutive polar
seasons from 2004 to 2011. To distinguish the data sets that have been acquired
throughout this construction phase, the different configurations are
denoted by the numbers of strings (e.g. IC59 for the 59-string
setup completed in 2010).

With a horizontal spacing of 125\un{m} and a vertical sensor spacing of 17\un{m}
on the majority of strings, the best sensitivity is achieved for
neutrino-induced muons above $\sim100\un{GeV}$. In the center of the array, the
denser spaced \deepcore array, yields a lower muon threshold of
$\sim10\un{GeV}$. Waveforms of the PMT's
pulse from the Cherenkov photons are digitally captured and timestamped with
$<3\un{ns}$ precision before being transmitted to the surface, where the data is
filtered and transmitted north via satellite. Using a likelihood method based on
detailed modeling of the optical scattering and absorption properties of the ice~\cite{icepaper},
the Cherenkov cone of the muon traversing the ice can be
reconstructed. This reconstructed direction is used as a primary
discriminator between the abundant flux of atmospheric muons from cosmic ray
air-showers from above the detector and muons induced by neutrinos interacting
in the ice or bedrock below the detector.
 
Despite the scattering of the Cherenkov light in the ice, good pointing can
be achieved due to the length of the muon track in the detector. The resulting angular resolution
for $\nu_\mu$ (including the $\nu$-$\mu$ scattering angle of $\left<\theta_{\nu\mu}\right> \approx
1\un{\deg}/\sqrt{E_\nu/\un{TeV}}$) obtained from simulation as achieved in a point-source
analysis with \icecube is shown in Figure~\ref{fig:angres}. Typically an accuracy of
$\sigma_\theta(\nu_\mu) \approx 1^\circ\un{deg}$ is achieved at $E_\nu = 1\un{TeV}$.
The accuracy of the resolution as well as the absolute pointing has been
verified by observation of the moon shadow in cosmic-ray induced muons~\cite{pointsource}, which
has now been observed with more than $13\sigma$.
The energy of the neutrino has to be estimated from the energy loss of the muon,
which is increasingly dominated by stochastic processes at higher energies, and
is thus limited to $\sigma_E(\mu) \sim 0.3 \log_{10}(E_\mu)$. The muon energy
yields only a very coarse proxy for the neutrino energy which is only partially
transferred to the muon (that is also typically not contained in the detector). For
$\nu_e$-induced cascades that are fully contained in the detector, a much better
energy resolution of $\sigma_E(\nu_e) \sim 0.13 \log_{10}(\nu_e)$ is achieved,
albeit at an order of magnitude lower effective area.

\twofig
{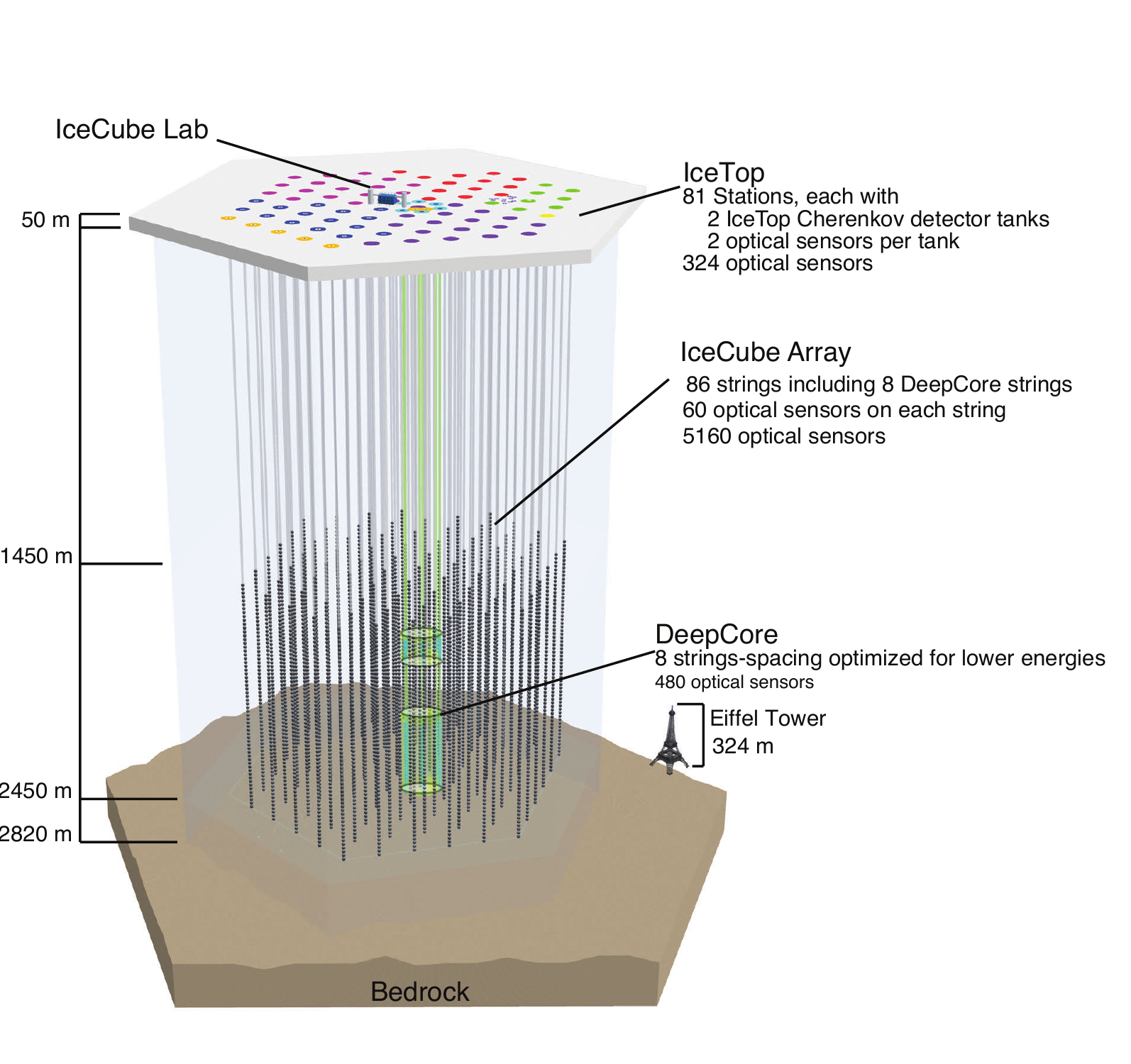}
{Schematic view of the \icecube detector with the subdetectors \deepcore and
\icetop.\label{fig:icecube}}
{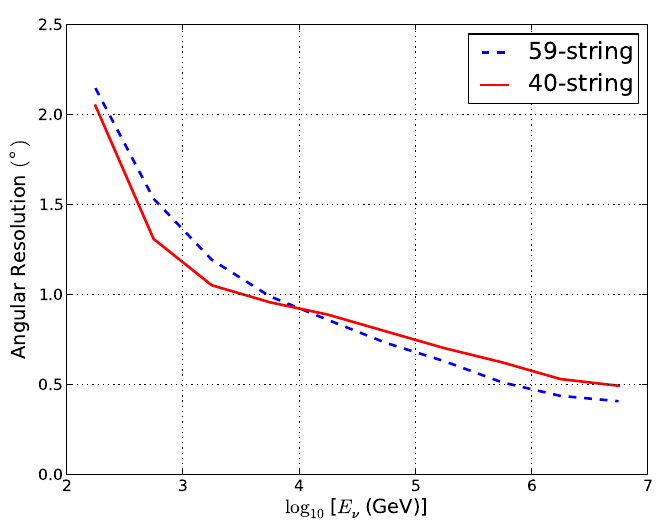}
{Median neutrino angular resolution for $\nu_\mu$ as a function of true neutrino energy
 for a cut-based (IC40) and boosted decision
tree-based (IC59) analysis~\protect\cite{pointsource}.\label{fig:angres}}

This \inice part of the observatory is augmented by a 1\un{km^2} air-shower
detector \icetop on top of the array. 324 of the same optical modules are placed in
pairs in ice filled tanks of 2\un{m} diameter, two of which are placed above every string.
This geometry provides a combined aperture of $A_\mathrm{eff}\cdot \Omega = 0.3
\un{km^2 sr}$ that yields about $10^7$ events per year above 300\un{TeV} coincidently detected in the
\inice and \icetop part of the array.

\section{Neutrinos}
The detection of extra-terrestrial high-energy neutrinos is the main objective
of \icecube. Through the interactions of protons with ambient matter $p+N \to
\pi^{\pm/0} + X$ or
radiations fields $p+\gamma \to \Delta^+ \to \pi^{\pm/0} + p/n$
and subsequent decay of the pions $\pi^0 \to \gamma\gamma$ and $\pi^{\pm} \to
\mu^\pm + \nu_\mu/\bar\nu_\mu \to e^\pm + \nu_e/\bar\nu_e +\nu_\mu/\bar\nu_\mu$
the sources of high-energy cosmic rays are intimately linked to high-energy
photons and neutrinos. 

\twofig
{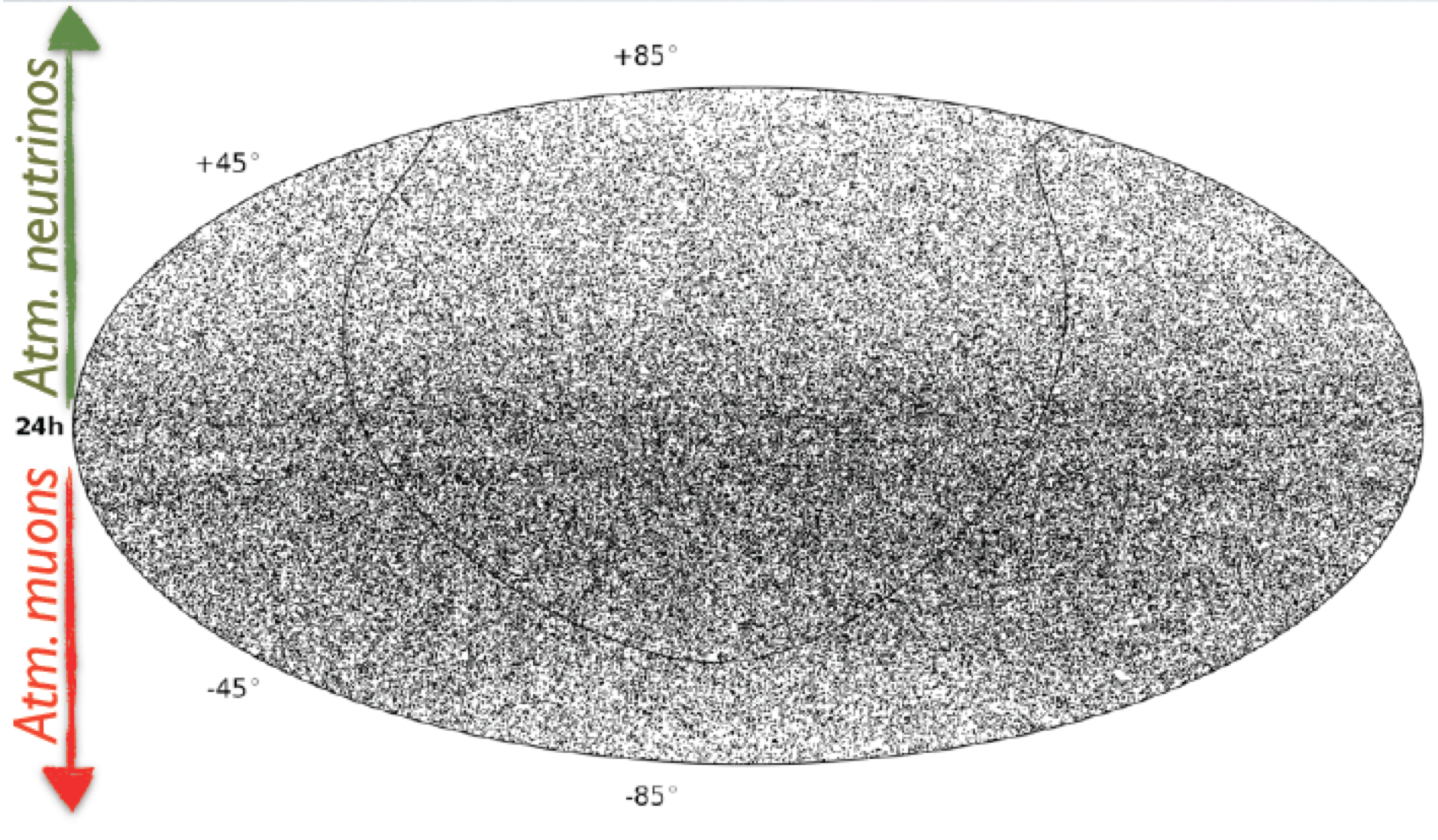}{Skymap of neutrino candidates in equatorial coordinates
for IC40+IC59. The curved
line indicates the galactic plane.
\label{fig:skymap}}
{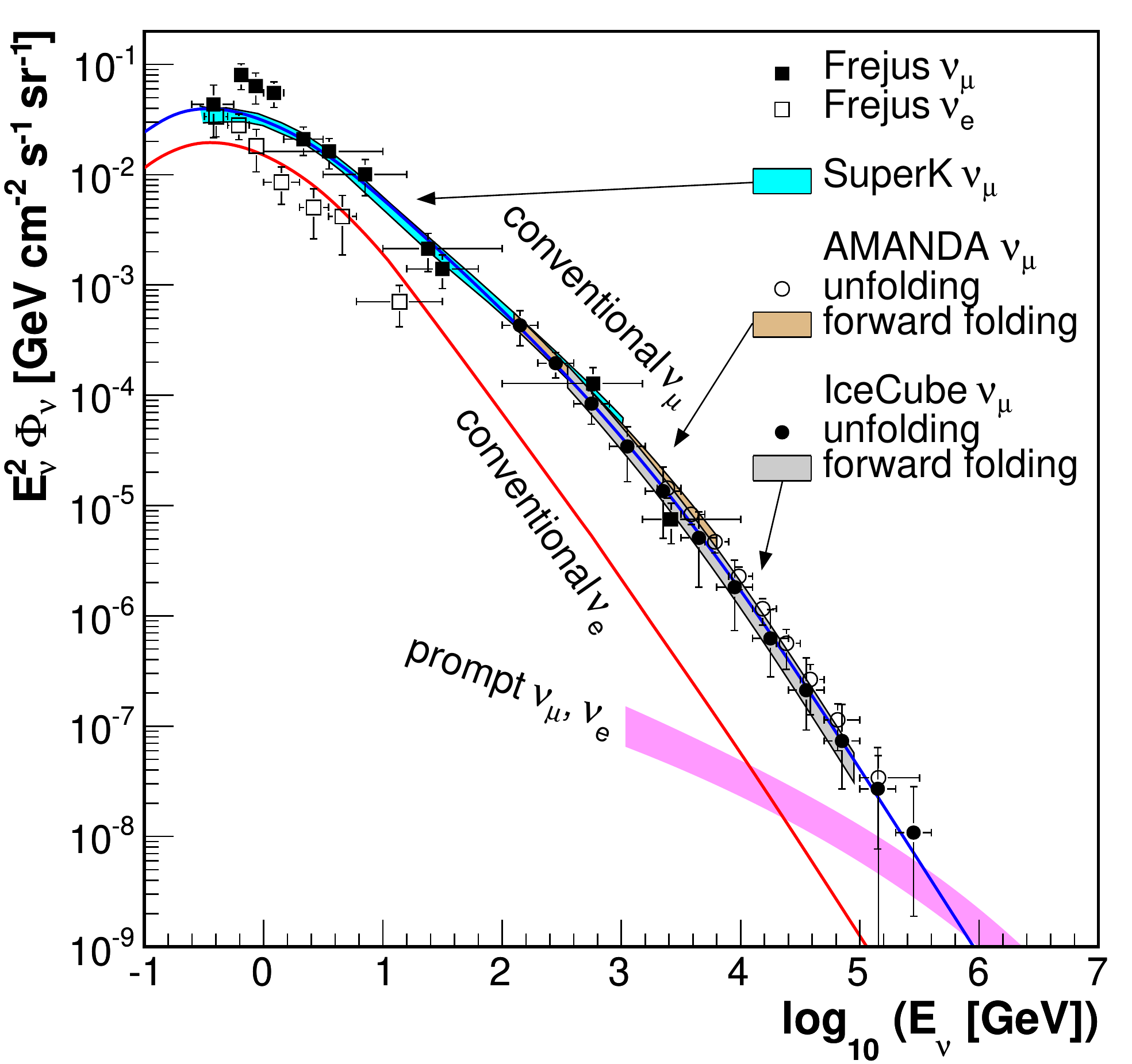}{Spectrum of atmospheric neutrinos from unfolding of the IC40 data
set~\protect\cite{diffuse}.\label{fig:atmnu}}
\subsection{Point Sources}
While charged cosmic rays are scattered in intergalactic magnetic
fields and photons are easily absorbed on the interstellar radiation fields or
the cosmic microwave background, neutrinos can free-stream from the sources, and
reveal their identity by a flux enhancment 
from that particular direction. Figure~\ref{fig:skymap} shows a skymap of
arrival directions of neutrino candidates. The plot contains
57460 up-going and 87009 down-going neutrino candidates
selected from 723 days of data taken with IC40 and IC59.
In the Northern sky it is dominated by atmospheric neutrinos.
As the flux of atmospheric muons falls with a much steeper spectral index of
-3.7 than anticipated from a neutrino source, this search can be extended to the
Southern sky by raising the energy threshold with zenith angle.

In a fine-binned full-sky likelihood analysis incorporating the number of likely
signal events as well as a spectral index for each point the hottest spot at
(Ra,Dec) = (75.45\un{\deg},8.15\un{\deg}) has a pre-trial p-value of
p=$2.23\cdot10^{-5}$ and a post-trial p-value of 0.67, indicating a high
compatibility with the null hypothesis~\cite{pointsource}. The resulting
zenith-dependant limit on neutrino fluxes from point sources not only exceeds
the initial sensitivity expected for the full detector configuration, but also
poses competitive limits in the Southern hemisphere. In addition, limits on the
neutrino emission are derived for a list of preselected source candidates~\cite{pointsource}.
No significant point source has been observed in either
search.

\subsection{Diffuse neutrino flux}
Following an analytic approximation~\cite{atmnu}, the flux of atmospheric neutrinos can be approximated as a sum
over the meson decay channels
$
\phi_\nu(E_\nu) = \phi_N(E_\nu) \sum_{\pi,K,D,\Lambda_c,\ldots}\frac
{A_i}{1+B_i\cos\theta E_\nu/\epsilon_\nu}
$
where $\phi_N(E_\nu)$ is the primary spectrum of nucleons evaluated at the energy of
the neutrino and $\theta$ is the zenith angle at the effective production
height. The most important contribution to 
at \un{TeV} energies is the
''conventional'' flux from $K^\pm \to \mu^\pm + \nu_\mu / \bar\nu_\mu$ and a
smaller contribution from $\pi^\pm \to \mu^\pm + \nu_\mu / \bar\nu_\mu$.  As
the neutrino energy increases above the critical energy
$\epsilon_i/\cos\theta$ ($\epsilon_\pi \sim 110\un{GeV},\epsilon_K
\sim 820\un{GeV}$), the spectrum steepens -- asymptotically by one power of
energy with respect to the primary spectrum.
Since the muon carries an unknown fraction of the
neutrino energy, a complex deconvolution mechanism has to be employed to infer the
spectrum. Figure~\ref{fig:atmnu} shows the result of an unfolding of the IC40
data of downgoing neutrinos. While the contribution of the
''prompt'' neutrino flux from the decays of charmed mesons is significantly
smaller in the $\un{TeV}$-regime, the critical energy is in the order of
$10^7\un{GeV}$. Hence their contribution will follow the primary
spectral index of the charged cosmic rays over the sensitivity range of
\icecube and they will become the dominated flux in the $\un{PeV}$-regime, which
is expected to be accessible with the full statistics of the completed detector.

\twofig
{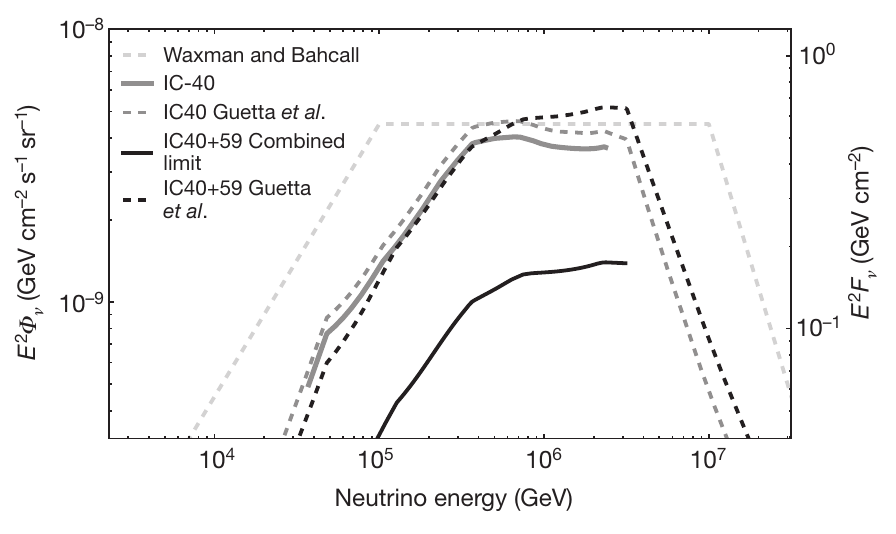}
{90\% C.L. upper limits (solid lines) on the neutrino fluxes from GRBs set IC40 and
IC59~\protect\cite{grbnat} with respect to the flux
expected from the model of Guetta et al.~\protect\cite{guetta} (dashed lines). 
The Waxman--Bahcall flux~\protect\cite{bahcall} assumes an average shape of the GRB
spectra.\label{fig:GRB}}
{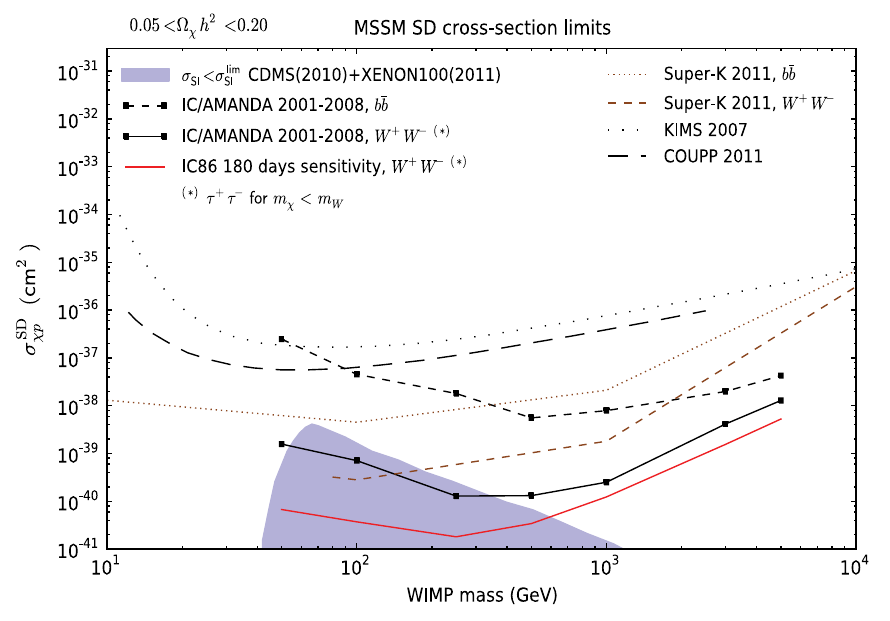}
{Limit on the \wimp-induced muon flux for annihilation to $b\bar{b}$ (black
dotted) and $W^+W^-$ (black) as well as the predicted senstivity for IC86 (red).
The shaded area shows the allowed MSSM parameter region with constraints from
CDMS and XENON~\protect\cite{wimp-sun}.\label{fig:wimps}}

\subsection{Gamma Ray Bursts}
In the fireball model, GRBs are modeled as explosions
of very massive stars which eventually collapse to a
black hole. In such models the observed gamma rays stem
from synchrotron radiation and/or inverse Compton scattering
of electrons accelerated in shock fronts in the collimated
explosive outflow. It was proposed that in the same
way also protons are accelerated~\cite{vietri,waxman}. These protons
would undergo interactions with the surrounding photon
field in the fireball and thus generate neutrinos~\cite{waxman,guetta}.

In a search for neutrinos from GRBs using the IC40 and IC59 detector
configurations, 224 GRBs reported by the GCN in the northern sky have been selected. A
likelihood was derived using a flat arrival time prior inbetween the first and
last photon observed for each GRB (typically $\Delta t = 0.1-100\un{s}$).
The estimated pointing resolution is taken into account for the neutrino
direction from \icecube as well as the GRB position. No neutrino event has been
observed against an expectation of 8.4 events from the prompt phase
model~\cite{guetta} with neutrino spectra calculated from the observed photon
spectrum for each GRB individually~\cite{becker}. Figure~\ref{fig:GRB} shows the
predicted neutrino flux as a function of energy for an average model~\cite{bahcall}
as well as for the specific set of GRBs entering the analysis,
together with the flux limit derived using the IC40, IC59 and combined IC40+IC59
dataset~\cite{grbnat}.

\subsection{\icecube{}'s Follow-up Program}
While \icecube continuously monitors the full northern hemisphere with
a typical uptime of above 99\%, most optical, X-ray and $\gamma$-ray
detectors have significantly smaller fields of view and/or duty-cycles.
In order to increase the coincident observation time, a number of
neutrino-triggered follow-up programs have been setup. In an online-analysis
chain, neutrino candidates are extracted from the data flow at the South Pole
with a latency of a $\sim6\un{min}$ and a purity of $>70\%$, from which alerts
are generated and forwarded to the northen hemisphere via a satellite connection.
\begin{itemize}
\item In the {\bf Optical follow-up} deviations from the atmospheric
neutrino flux are identified by searching for multiplets within a time
window of $\Delta t < 100\un{s}$ and angular distance $\Psi < 3.5\un{\deg}$.
Based on a likelihood criterion incoorperating the temporal and spatial distance
as well as the angular resolution, alerts are forwarded to the optical
telescopes ROTSE
and PTF.
A first analysis~\cite{ofupaper} using data from Dec. 2008 to Dec. 2009 yields
an insignificant $2.1\sigma$-excess of multiplets. No supernova has been
identified in 17 follow-up observations with ROTSE. For the first time a
dedicated search set stringent limits on models for choked jets in core-collapse supernovae~\cite{chokedjets}.
Since Mar. 2011, the program has been extended by
an {\bf X-ray follow-up} with the \swift satellite, aiminig at the observation
of the rapidly decaying afterglow from GRBs.
\pagebreak
\item For the {\bf $\gamma$-ray follow-up}~\cite{pointsource}
a set of 109 source candidates (predominantly AGNs) is identified
a-priory. A time-clustering algorithm searches for excesses from these
directions in the online neutrino stream within time scales of up to 3 weeks, taking into account the
temporal variability of the background.
Alerts from this program will be sent to the \magic and \veritas $\gamma$-ray
telescopes with an expectation of
one background-induced alert/year.
\end{itemize}

\subsection{\wimp searches}
It is speculated that \wimp{}s can lose energy through elastic scattering with
ordinary matter, and become trapped in gravitational wells.
The accumulated excess gives rise to an increased annihilation
rate, resulting in turn in a neutrino flux enhancement from decay of the
self-annhilation products.
\icecube is hence looking for neutrinos from \wimp
annihilation by searching for enhanced signals from the Earth's center~\cite{wimp-earth}, the
Sun~\cite{wimp-sun} or the galactic halo~\cite{wimp-gal}.
By assuming that in the Sun an equilibrium between capture and annihilation is
reached,
the neutrino-induced muon flux can be related to the
spin-dependant cross-section $\sigma_\mathrm{SD}$. In Figure~\ref{fig:wimps}
limits for $\sigma_\mathrm{SD}$ are given for \wimp masses
$m_\chi$ from 50\un{GeV} to 5\un{TeV}~\cite{wimp-gal}. As the
effective area of \icecube increases with energy, its sensitivity will depend
on the spectrum of the decay products.
Hence significantly stronger constraints can be placed on the decay $\chi\chi
\to W\bar{W}$ compared to  $\chi\chi \to b\bar{b}$.  This analysis
combines data taken with \icecube and the precursor detector
\ama between 2001 and 2008, with a total livetime
of 1065 days when the Sun was below the horizon~\cite{wimp-gal}.

The shaded area in Figure~\ref{fig:wimps} indicates the region
not yet excluded by the MSSM parameter constraints through the direct searches
by the experiments CDMS and XENON100. Comparing with the expected sensitivity
for a livetime of 180\un{days} for the full detector also shown in Figure~\ref{fig:wimps}
illustrates the potential of the approach.

\twofig
{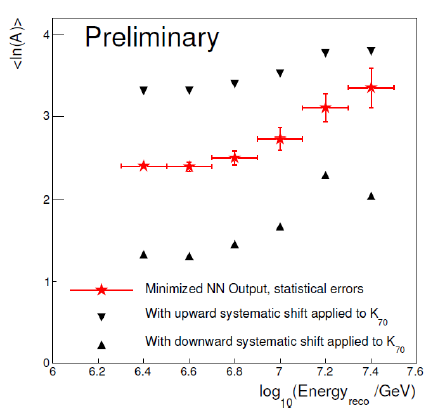}
{Average logarithmic mass of cosmic rays from one month of data with
IC40/IT40~\protect\cite{icetop}.\label{fig:mass}}
{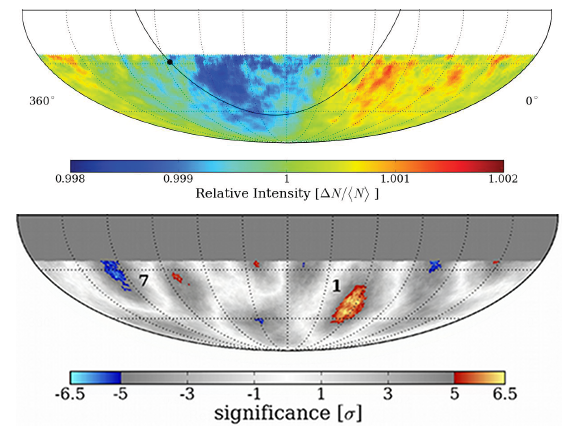}
{Relative intensity map for cosmic rays of
the 20\un{TeV} sample (top) and excess significance after subtraction of the
dipole and quadrupol component (bottom)~\protect\cite{icetop}.\label{fig:anisotropy}}

\section{Cosmic Rays}
\subsection{Spectrum and composition}
For measuring cosmic rays, \icecube is complemented by the
surface air shower array \icetop, which samples the electromagnetic shower
component, whereas the deep detector responds to punch-through muons from the
hadronic shower component. This combination
offers a unique possibility to determine the spectrum and mass
composition of cosmic rays from about 300\un{TeV} to 1\un{EeV}. The first
analysis exploiting this 
correlation~\cite{icetop} uses a small data set corresponding to only one month of data
taken with about a quarter of the final detector. A neural network was employed to simultaneously
determine the primary energy and mass from the measured input variables shower
size and muon energy. The resulting average logarithmic mass shown in
Figure~\ref{fig:mass} indicates that the average mass of the primary particles
increases with energy in the knee region around
$10^{15}\un{eV}$~\cite{icetop}.

\subsection{Cosmic Ray anisotropy}
A large sample of $10^{11}$ cosmic ray muon events has been collected by
\icecube between 2007 and 2010 (and will roughly increase by the same amount
every year with the full detector). For the first time, the anisotropies
previously reported on multiple angular scales~\cite{milagro,tibet}
could be studied in the Southern sky as well~\cite{aniso1,aniso2}.
In a multipole analysis, a dominant dipole contribution is found that does not
coincide with the direction of the Compton-Getting effect~\cite{compton}, indicating that the cosmic
rays co-rotate with the local galactic magentic field. Additional structures are
found on angular scales from $15-30\un{\deg}$ (c.f. Figure~\ref{fig:anisotropy}).
These are potentially ascribed to
the local magnetic field and source configuration, and may hence carry key
information on the origin of galactic cosmic rays.

\subsection{Future enhancements}
It has been shown recently that air-showers can also be detected by the coherent
radio emission in the $10-100\un{MHz}$ regime~\cite{lopes-nat}
dominantly stemming from deflection of the electrons in the earth magnetic field and a net charge
excess in the shower~\cite{reas,mgmr}. In contrast to \icetop which samples the
shower on the ground, the basically unattenuated radio signal reflects the
integral development of the shower through the atmosphere, and might
additionally yield an independant handle on the shower maximum~\cite{mgmr-comp}.
It has hence been suggested~\cite{rasta} that complementing the \icecube
observatory with an array of radio receivers could contribute significantly to a
precise measurements of the cosmic ray spectrum and composition. Results from a
first test setup~\cite{ara-perf} indicate very low ambient noise levels at the
South Pole and therefore the good suitability of the location for this approach.

\section{Conclusion}
While no extra-terrestrial sources of neutrinos have been identified yet, the
physics results obtained with \icecube throughout the construction period
demonstrate the excellent performance. The increased statistics available with the
now complete detector and continous improvements in reconstruction and
background rejection propose \icecube will continue to make key contributions to
a wide class of physics cases in the future.
\\

\begin{multicols}{2}[\section*{References}]
\vspace*{-2\baselineskip}

\end{multicols}

\end{document}